\documentclass[twocolumn]{revtex4}
\usepackage{graphicx}
\begin{document}
\draft
\title{DETERMINISTIC RATCHETS: ROUTE TO DIFFUSIVE TRANSPORT}
\author{M. Borromeo}
\affiliation{Dipartimento di Fisica, and Istituto Nazionale di Fisica Nucleare, Universit\'a di Perugia, I-06123 Perugia (Italy)}
\author{G. Costantini}
\author{F. Marchesoni}
\affiliation{Istituto Nazionale di Fisica della Materia, Universit\'a di Camerino, I-62032 Camerino (Italy)}
\date{Received:\today}
\begin{abstract}
The rectification efficiency of an underdamped ratchet operated in the {\it adiabatic} regime 
 increases according to a scaling current-amplitude curve as
the damping constant approaches a critical threshold; below threshold the rectified signal becomes
extremely irregular and eventually its time average drops to zero. 
Periodic (locked) and diffusive  (fully chaotic) trajectories
coexist on fine tuning the amplitude of the input signal. The transition from regular to chaotic 
transport in noiseless ratchets is studied numerically.
\\

PCAS numbers: 05.60.Cd, 05.45.-a, 87.16.Uv\\
\end{abstract}
\maketitle

\section{INTRODUCTION}

Particles in an asymmetric potential can drift on average in one direction even when the time and 
space average of all applied macroscopic forces or gradients is zero. To achieve directed
transport in such a device, called a ratchet, a time-correlated source of energy is required,
for instance non-markovian external fluctuations (thermal ratchets \cite{ratchets}) or a time 
periodic drive (rocked ratchets \cite{bart}). In signal analysis notation, a ratchet can be regarded 
as a (random or periodic) signal rectifier. Rocked ratchets with massive particles exhibit strong
inertial effects capable of reversing their current \cite{inertia}. Moreover, inertial 
rocked ratchets are naturally subject to developing chaotic dynamics, the onset of which becomes
detectable as thermal fluctuations are switched off \cite{jung,mateos}.

This class of devices, operated at zero noise level, may be assimilated to an asymmetric version
of the damped-driven pendulum, a chaotic system  investigated at depth in the 
Eighties \cite{huber}-\cite{gollub}. The dynamics of a massive particle in a cosine potential
was reproduced in terms of a "climbing-sine map" \cite{kapral}: Running orbits,
relevant to the rectification mechanism in a ratchet, can be either periodic or diffusive, depending
on the value of the map control parameter (viz. the amplitude of the sine term). The phase-space
portrait of the actual damped-driven pendulum was computed by Huberman {\it et al.} \cite{huber},
who revealed the existence of {\it delocalized} strange attractors with an intricate structure on
all scale, later recognized to be fractal objects \cite{gollub}.

In this article we investigate the transport of unit mass particles in a ratchet potential subjected
to a sinusoidal driving force $A(t)$ with angular frequency $\Omega$ much smaller  than the damping 
constant $\gamma $ and the librational frequency $\omega_0$ of a particle trapped at the bottom of a 
potential well. The damping constant is lowered from exceedingly large values $\gamma > 10\omega_0$,
where the Smoluchowski approximation of Ref. \cite{ratchets} holds good, deep into the 
underdamped regime $\gamma \sim 0.01\omega_0$. The adiabatic condition $\Omega < \omega_0, \gamma$
imposed here, has not been addressed in the damped-driven pendulum literature, but is believed 
 to maximize transport in biological systems \cite{ratchets} and Josephson junction
arrays \cite{huber,squids,gold}, among others. On decreasing $\gamma$ we noticed that (a) The rectification
efficiency of an inertial ratchet increases inversely proportional to $\gamma$
according to a scaling current-amplitude $\langle v \rangle$-$A_0$ curve; 
(b) Below a critical  $\gamma$ value,
more $\langle v \rangle$-$A_0$ 
curves coexist, thus signaling the appearance of regular trajectories with higher periodicity;
(c) On further
decreasing $\gamma$, all trajectories develop a finite {\it diffusion} coefficient
and the ratchet current drops monotonically toward zero.
These results are of {\it immediate} application to the design and
operation of nano-array devices, both in solid state- \cite{squids,gold}
and in bio-engineering \cite{ratchets}.

\section{THE DOUBLE-SINE MODEL}

We have integrated numerically the archetypal ratchet model \cite{bart}
\begin{equation}
\label{1}
\ddot x=-\gamma \dot x -V'(x) + A_0\sin(\Omega t),
\end{equation}
with asymmetric double-sine potential (Fig. \ref{F0})
\begin{equation}
\label{2}
V(x)=-{1\over k}[\sin(kx) +{1\over 4}\sin(2kx)]
\end{equation}
and arbitrary time origin of the drive $A(t)=A_0\sin(\Omega t)$. The constant $k$ was set equal
to $\pi/2$, so that the unit cell of (\ref{2}) is $l=4$ and the libration frequency at the bottom
of its wells is $\omega_0=\sqrt{k(3\sqrt{3}/2)^{1/2}}\simeq 1.591$. The rectification window $(A_1,A_2)$
of the ratchet (\ref{1})-(\ref{2}) is determined by the  depinning thresholds of the tilted
potentials $V(x)\mp x|A_0|$, i.e. $A_1=3/4$ to the right (-) and $A_2=3/2$ to the left (+).
As $A_1<A_2$, the ratchet current is {\it naturally} directed to the right \cite{bart} or,
equivalently, the time average $\langle v \rangle$ of the particle velocity is positive definite
for $A_0>A_1$.  
The forcing frequency $\Omega$ is taken so small that the the $\Omega$-dependence of $A_1, A_2$
is negligible \cite{jung}. 

In the adiabatic limit, $\Omega \ll \omega_0, \gamma$, the values of
$\langle v \rangle$ are quantized with step $\Delta v=l/T_{\Omega}$, corresponding to a net particle
shift of one unit cell $l$ per forcing period $T_{\Omega}=2\pi/\Omega$. Of course, quantization
 of the $\langle v \rangle$-$A_0$ characteristics becomes less and less apparent as the 
ratchet operating conditions most relevant to the present investigation, i.e. $\Omega \ll \gamma \ll
\omega_0$, are approached (see Fig. \ref{F1}). The integration of Eq. (\ref{1}) was performed
by means of a 4-th order Runge-Kutta algorithm
with integration step $\Delta t=2\cdot10^{-6}T_{\Omega}$.

In our numerical work we set $\Omega=0.01$ and explored the ratchet response on varying the damping
constant in the interval $[10^{-2}, 10]$. For large $\gamma$ values ($\gamma =10$ in Fig. \ref{F1})
we reproduced the results of Ref. \cite{bart}. The current-amplitude curve grows linearly 
with $A_0$ from zero 
up to a maximum $(A_2-A_1)^2/2$, between $A_1$ and $A_2$, and then decays with comb-like structure
reminiscent of the Vernier effect. 
[The amplitude of the particle excursions both to the left and to
the right increases fast with $A_0$, but slightly faster to the left
than to the right.] Note that 
the rectification effect persists for values of $A_0$ much larger than $A_2$.

\subsection{The Mobility curves}

For $\gamma$ values in the range $[0.07, 0.5]$, the $\langle v \rangle$-$A_0$ characteristics exhibits
the universal behavior displayed in Fig. \ref{F1}. The data points of $\gamma \langle v \rangle$ 
versus $A_0$ rest on a unique limiting curve ($\gamma= 0.1$ in Fig. \ref{F1}) with deviations of a 
few percent (and diminishing with lowering $\gamma$). The scaling law $\langle v \rangle \propto 1/\gamma$ 
for an adiabatically driven underdamped particle is to be expected \cite{risken,io}. 
More remarkably, we observed that: 

(i) The ratchet current gets more and more narrowly restricted
to the rectification window $(A_1,A_2)$ as the damping constant grows smaller than $\omega_0$. 
This effect is even more apparent in Fig. \ref{F2} for $\gamma=0.05$; 

(ii) The onset of the ratchet
current at the threshold $A_1$ is abrupt; $\gamma \langle v \rangle$ jumps from zero up to
0.086 within one incremental step $\Delta A_0=2\cdot 10^{-6}$. On the contrary, the sharply decaying
branch of the curve $\gamma \langle v \rangle$ for $A_0>A_2$ is
continuous, its persistent irregularity being a signature of the
underlying chaotic dynamics of the system (see caption of
Fig. \ref{F2}); peaks and dips
have been resolved on the scale $\Delta A_0=10^{-5}$, i.e. within the 
thickness of the broken line connecting our data points.

The universal $\gamma \langle v \rangle$-$A_0$ curve can be determined as follows. For 
$A_1<A_0<A_2$ the particle runs to the right with velocity $v(t)=A_0\sin(\Omega t+\phi)/\gamma$,
where the adiabatic phase-lag is $\phi=-\arctan(\Omega/\gamma)$. As discussed in Ref. \cite{risken},
a trapped particle gets depinned to the right when $A(t)> A_1$ and, then, repinned as soon as
$A(t) \leq A_r$, with $A_r\propto A_1(\gamma/\omega_0) \simeq 0$. An underdamped particle will be then
traveling in the positive direction during the time interval $[\Omega^{-1}\arcsin(A_1/A_0), \pi/\Omega]$
\, mod$(2\pi/\Omega)$; hence, the time average of $v(t)$ over one forcing period yields
\begin{equation}
\label{3}
{{\gamma\langle v \rangle}\over {A_0}}\equiv \mu(A_0,A_1)= {{1}\over{2\pi}}
[\sqrt{1 -{({{A_1}/{A_0}})^2}}+1].
\end{equation}
The r.h.s. of Eq. (\ref{3}) is $\gamma$ independent and accurate in the limit $\gamma/\omega_0 
\rightarrow 0$. For $A_0>A_2$ a similar argument leads to $\gamma \langle v \rangle/A_0=\mu(A_0,A_1)
-\mu(A_0,A_2)$ (see Fig. \ref{F1}, inset). The analogy with the experimental $I-V$ characteristics
of the Josephson ratchet in Ref. \cite{gold} is apparent.

A totally different scenario emerges on decreasing $\gamma$ below 0.07. The current for $A_0>A_2$ drops
close to zero, while in the rectification window the data points taken at steps of $\Delta A_0=10^{-3}$ arrange
themselves along different current-amplitude characteristics. In particular: 

(iii) In Fig.
\ref{F2}, besides the universal curve (\ref{3}) denoted by $1:1$, at least three more characteristics,
$1:2 - 1:4$, are resolved. In our notation curve $1:m$ corresponds to curve (\ref{3}) divided by
the integer $m$. Note that, contrary to the case of the damped-driven pendulum \cite{huber,kapral},
 in the asymmetric problem (\ref{1})-(\ref{2}) odd $m$ values are not ruled out; 

(iv) A 
$\gamma \langle v \rangle$-$A_0$ background is enclosed by curve $1:4$. 
(or, possibly, by $1:5$, not shown)
The data point distributions on the characteristics $1:1-1:4$ are denser
in the vicinity of $A_1$, while 
the background gets thicker close to $A_2$. 
[Sparse negative mobility background values have been recorded immediately above the
threshold $A_1$.]
In the window $(A_1,A_2)$ compact $A_0$ intervals corresponding to
the diverse characteristics (or to the background) can be easily 
indentified on the scale
$\Delta A_0=10^{-5}$ (see Fig. \ref{F2}, inset). 
On further decreasing $\gamma$ below 0.04, the curves
$1:1-1:4$ get depleted and only an extremely irregular, low background survives. 

In order to explain the coexistence of many $\langle v \rangle$-$A_0$  characteristics, we sampled one 
trajectory from each of the uppermost three $1:m$ curves (Fig. \ref{F3}, top panel, upper). These trajectories are
seemingly periodic and independent of the initial conditions. The trajectory corresponding to curve
$1:1$ drifts with positive speed stepwise, by alternating one plateau (trapped particle) with one jump 
of constant length $\Delta x_p$ (running particle). The trajectory representative of curve $1:2$
consists of a regular sequence where each plateau is followed by a full oscillation (here, 
with measured peak-to-peak amplitude  $L\simeq3800$).
Note that the separation $\Delta x_p$ between two adjacent plateaus is almost the same as in the first case,
being the $A_0$ difference as small as $10^{-4}$.
Consequently, the average velocity of a type $1:2$ trajectory is half the reference value (\ref{3})
on curve $1:1$. Analogously, a type $1:3$ trajectory alternates periodically one plateau each
two oscillations, thus drifting with one third of the velocity (\ref{3}).

\subsection{The Diffusion Coefficient}

The behavior of the background trajectories is even more interesting: 

(v) Plateaus are separated by {\it random}
sequences of full oscillations, extremely sensitive to the initial conditions
(Fig. \ref{F3}, top panel, lower); the lengths $\tau$ of such sequences
are distributed according to a simple {\it exponentially} decaying
law. Occasionally, the oscillation sequences slant
to the left, i.e. in the negative direction, with glitches of different length. As suggested in
 Refs. \cite{huber,jung}, we determined the second cumulant of the {\it random} trajectories plotted in Fig.
\ref{F3}, 
\begin{equation}
\label{3.1}
C(t)= \overline{x^2(t)}-\overline{x(t)}^2,
\end{equation}
the averages $\overline{(\dots)}$ being taken
over the particle initial conditions $x(0), \dot x(0)$ \cite{jung}. From the fitting law 
\begin{equation}
\label{3.3}
\lim_{t\rightarrow\infty}C(t)=2Dt,
\end{equation}
one eventually extracts the diffusion coefficient $D$ of the noiseless
process $x(t)$ (Fig. \ref{F3}, bottom panel);

(vi) Trajectories change from regular to chaotic  on tuning $\gamma$ at constant $A_0$.
Typically, for values of $A_0$ corresponding to points sitting on curves $1:2-1:3$ in Fig. \ref{F2},
the periodicity of the $x(t)$ trajectories varies as $\gamma$ is ramped from 0.07 
down to 0.04; trajectories with different periodicity show up repeatedly, interspersed with diffusive trajectories; eventually,
the diffusive nature of the underdamped processes prevails. 
Reentrant type $1:1$ trajectories
do not imply the current reversals observed in Ref. \cite{mateos}. 
For $\Omega=0.01$, type $1:m$ trajectories with $m\geq 4$ 
are hardly detectable; moreover,
these different behaviors turned out to be insensitive to the initial conditions.

The onset of diffusive transport is a main feature of inertial ratchets \cite{jung}. Although an analytical
estimate of the diffusion coefficient $D$ lies beyond the grasp of our theoretical tools  
a simple relationship between drift and diffusion of a ratchet (\ref{1})-(\ref{2})
in the fully chaotic regime can be easily established.
 Consider the case $A_0=0.9305$ represented in Figs. \ref{F2} and \ref{F3}.
The average value of $\gamma \langle v \rangle$ is 0.045, namely one fifth of the corresponding value on 
curve $1:1$; thus, the average duration of the oscillation sequences along the relevant diffusive trajectories is
of the order $\overline{\tau}=nT_{\Omega}$ with $n=4$. The separation $\Delta x_p$ between two subsequent plateaus 
[no matter what $n$, as long as one
disregards the negative slant of the oscillation sequences] can be evaluated by integrating $v(t)=
A_0\sin(\Omega t+\phi)/\gamma$ along the running solution branch, i.e.  over the time interval
$[\Omega^{-1}\arcsin(A_1/A_0), 
\pi/\Omega]$; for the parameter values of Fig. \ref{F3}, $\Delta x_p\simeq (3/4)L$. As a consequence, the standard formula \cite{cox}
\begin{equation}
\label{3.2}
D=(\Delta x_p)^2{{\overline{\tau^2}-\overline{\tau^{\,}}^2}\over{2\overline{\tau^3}}}
\end{equation}
 yields the estimate $D
=(3L/4)^2/12\overline{\tau}\simeq 270$
that compares fairly closely with the fitted value $D=380\pm 50$ 
(in the units of Fig. \ref{F3}).

\section{PHASE-SPACE PORTRAITS}

The chaotic nature of the trajectories in Fig. \ref{F3} is illustrated by the reduced phase-space portrait
of Fig. \ref{F4}; here, $\gamma v$ and $\overline{x} \equiv x \,\mbox{mod}(l)-(1/2)$ have been recorded
once every forcing cycle at $t_n=(n+1/4)T_{\Omega}$ (stroboscopic snapshots). The periodicity of
type $1:1-1:3$ trajectories is apparent: In all three cases there exists one single point at a lowest
velocity ${v_0}$, associated to
the depinning of the ratcheted particle out of the potential trap. Type $1:2$ and $1:3$ trajectories exhibit one
or two {\it full} oscillations, respectively, prior to repinning (Fig. \ref{F3}); therefore,
the snapshot velocities corresponding to such
oscillations are larger than the relevant ${v_0}$. In view of a straightforward energetic argument, one concludes 
that the portrait points representative of regular trajectory oscillations must be located along the upper
asymmetric band of Fig. \ref{F4}, $v_1(\overline{x})$, defined by
\begin{equation}
\label{4}
v_i(\overline{x})-\overline{v_i}= - V(\overline{x}-\overline {x_0})/\overline{v_i},
\end{equation}
where $i=1,2$ is a band index, $\overline{v_i}$ its mean value, $V(x)$ the ratchet potential (\ref{2}) and
$\overline{x_0}$ a spatial shift that depends on the definition  of $\overline{x}$ and on the velocity
phase-lag $\phi$.

Most remarkably, the phase-space portrait representative of regular trajectories consists of singular points (more likely, confined segments),
denoted by 1--3 in Fig. \ref{F4}; this implies that such trajectories are truly periodic and that,
equivalently, for appropriate parameters values inertial depinning amounts to a
{\it renewal} process \cite{cox}
of variable locking time. The velocity gap $\overline{v_1}-\overline{v_0}$ in our stroboscopic rappresentation
can be easily estimated. Let us assume that, on neglecting a small phase-lag $\phi$, the particle velocity
is zero at $t=0$ (no tilt) and relaxes towards its asymptotic value $A(t_1)/\gamma=A_0/\gamma$ according to
an exponential law, namely, ${v_0}=(A_0/\gamma)[1-e^{-\gamma(t_1-t_0)}]$, if the particle exits a potential 
well assisted by the tilt $A(t)$ at $t_0=\Omega^{-1}\arcsin(A_1/A_0)$, or $\overline{v_1}=(A_0/\gamma)
[1-e^{-\gamma t_1}]$, if, thanks to inertia, it executes a full oscillation without getting trapped. The ensuing
estimate $\gamma (\overline{v_1}-{v_0}) \simeq 3.8 \cdot 10^{-2}$ agrees fairly well with our data
in Fig. \ref{F4}.

The reduced phase-space portrait of a diffusive trajectory is characterized by a confined segment ("a" in Fig. \ref{F4}),
associated with the depinning events, and a complicated gapped band structure, accounting for the negative 
trajectory slant. The splitting of the upper edge into two close bands $v_i(\overline{x})$ with $i=1,2$ is 
related to the existence of two flexural points on the shallow, rising branches of $V(x)$. The dispersion
of the representative points along the bands $v_i(\overline{x})$ and the bottom ribbons
(same shape as $v_i(\overline{x})$, but apparently quantized along the $\gamma v$ axis) is a clear-cut
 signature of a {\it fully chaotic} behavior. 

\section{CONCLUSIONS}

In this report we focused on certain properties of {\it adiabatic
 inertial ratchets} that are of direct application to bio-engineered and nanoelectronic
devices. Related questions that might deserve closer attention by
 theorists and experimenters, alike, are:
(a) The  robustness (or selectivity) of curves $1:m$ 
in the presence of noise. In particular,
one wonders if all characteristics $1:m$ may survive in the
presence of thermal fluctuations. Most likely, with increasing
the temperature, the diffusive nature of the ratchet trajectories
will prevail in the vicinity on $A_2$, first, and eventually over the entire
rectification window.
(b) The dependence of the diffusion coefficient on the
ratchet parameters.
In the absence of noise, the diffusion coefficient
depends crucially on the nature of the relevant trajectories.
Trajectories may be periodic ($D=0$), fully chaotic (when the
argument of Eq. (\ref{3.2}) applies) or exhibit an intermediate
behavior, termed {\it confined chaos}. For trajectories in this
category all singular points in the phase-space portraits of
Fig. \ref{F4} are replaced by curves of different length and the diffusion
coefficient acquires a finite value.
(c) The generalization of the present analysis to wider classes of asymmetric
potentials \cite{io,asym}.  While the mobility curves may depend on the
choice of $V(x)$, the dynamical regime $\Omega \ll \gamma \ll \omega_0$
is expected to show important similarities, no matter what the details of
the substrate model.

\newpage

\begin{widetext}

\begin{figure}
\centering
\includegraphics[scale=0.7, angle=270]{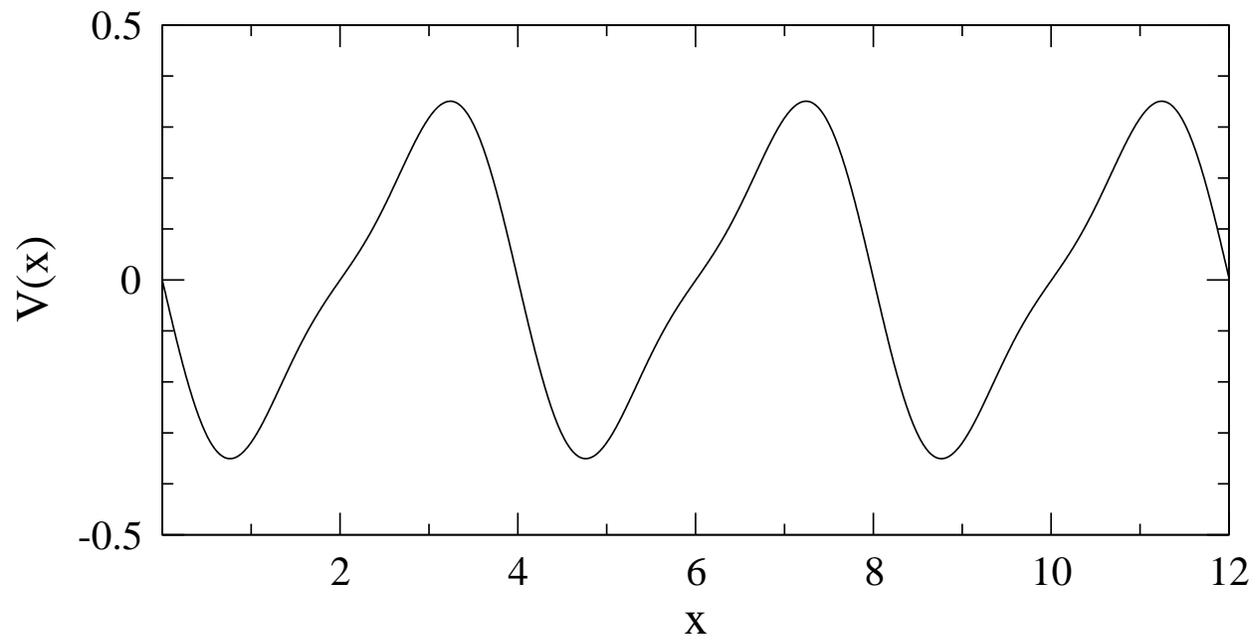}
\caption
{\label{F0}
The double-sine potential (\ref{2}) with $k=\pi/2$. As the l.h.s. of each potential
well is steeper than the r.h.s., the natural ratchet direction is positive.
}
\end{figure}

\begin{figure}[t]
\includegraphics[scale=0.7, angle=270]{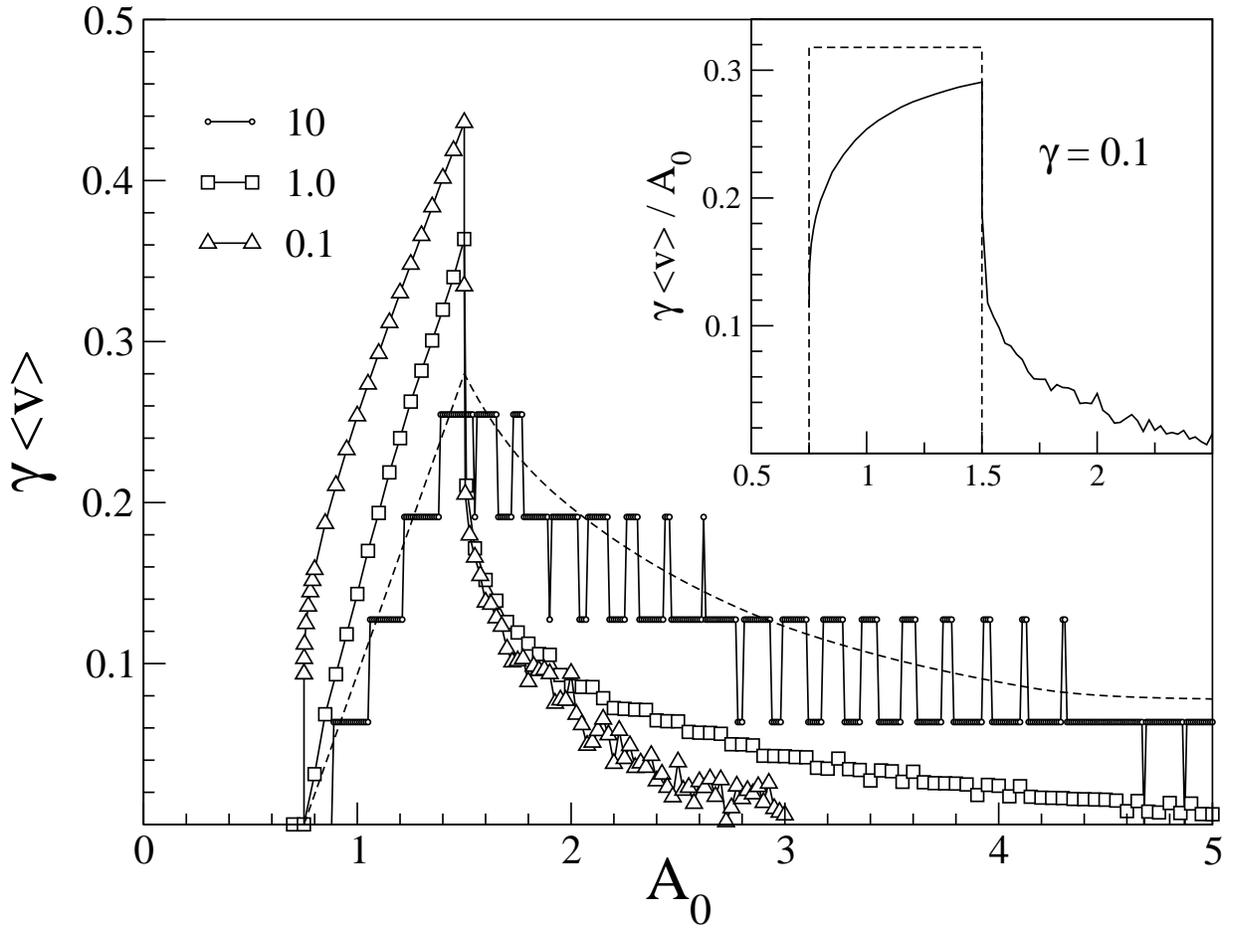}
\caption{\label{F1}
$\gamma \langle v \rangle$-$A_0$ characteristics of ratchet (\ref{1})-(\ref{2}) for $\Omega=0.01$
and different values of $\gamma$ (incremental steps $\Delta A_0=10^{-3}$). The dashed curve represents
the adiabatic limit $\Omega=10^{-4}$ and $\gamma = 10$. Inset: the numerical data for $\gamma=0.1$
(solid curve) are compared with $\mu(A_0,A_1)$ for $A_1 < A_0 < A_2$, Eq. (\ref{3}),
and  $\mu(A_0,A_1) - \mu(A_0,A_2)$ for $A_0 > A_2$ (dashed curve). The velocity averages have
been taken over 300 forcing cycles, after discarding the first $10^3$ cycles to get rid of
transient effects.
}
\end{figure}

\begin{figure}[t]
\includegraphics[scale=0.7, angle=270]{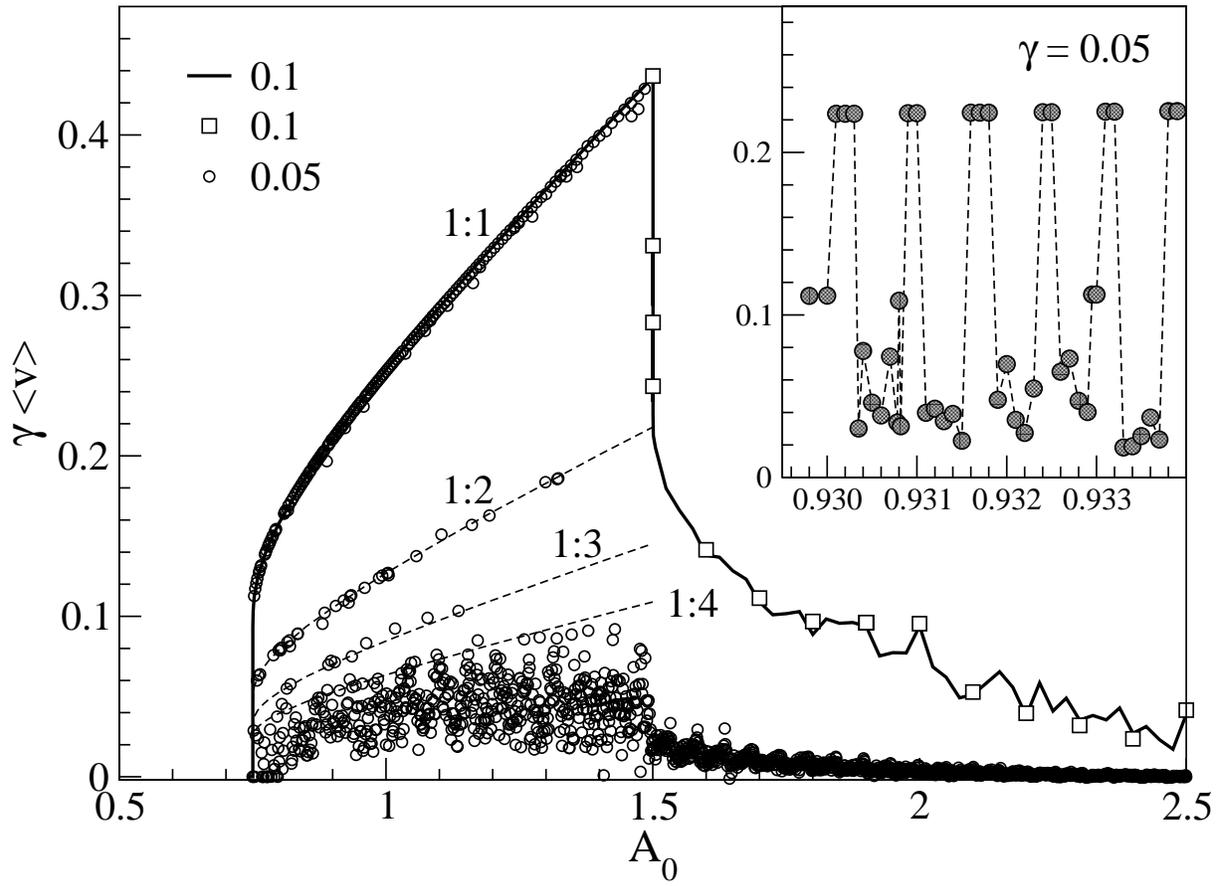}
\caption{\label{F2}
$\gamma \langle v \rangle$-$A_0$ characteristics of ratchet (\ref{1})-(\ref{2}) for $\Omega=0.01$,
$\gamma=0.05$ and  $\Delta A_0=10^{-3}$ (circles). Inset: details with $\Delta A_0=10^{-4}$.
The curve $\gamma=0.1$ from Fig. \ref{F1} is plotted for reader's convenience (solid curve).
Averages have been taken as in Fig. \ref{F1}, but for the squares on curve $\gamma=0.1$, where
the preparation time was doubled ($2 \cdot 10^3$ cycles). Dashed curves: characteristics $1:1$-$1:4$.
}
\end{figure}

\begin{figure}[t]
\includegraphics[angle=270]{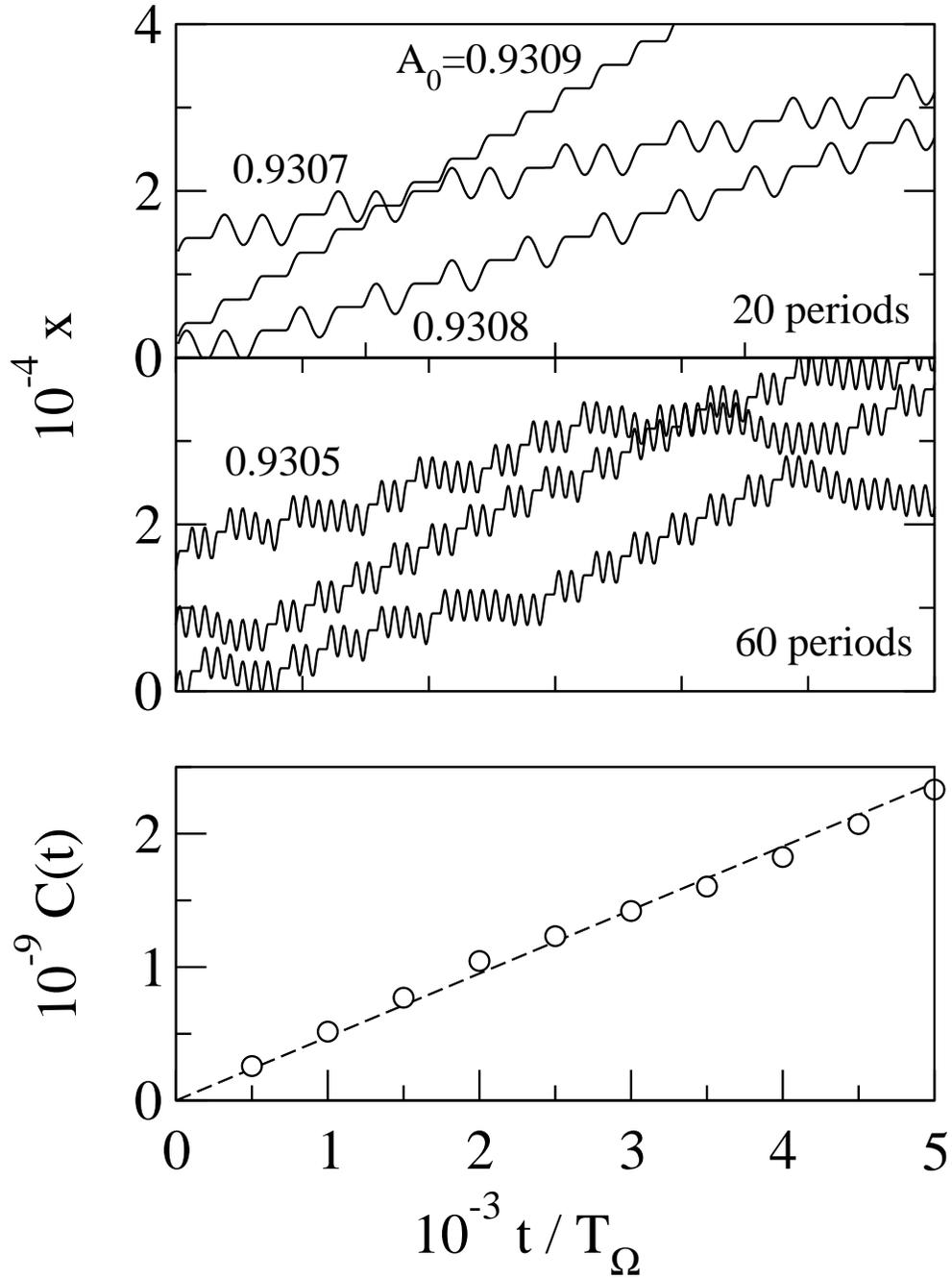}
\caption{\label{F3}
Top panels: samples of regular and diffusive trajectories (the latter ones for different initial conditions,
see text) for $\Omega=0.01$ and
$\gamma=0.05$; bottom panel:
 $C(t)= \overline{x^2(t)}-\overline{x(t)}^2$
 for $A_0=0.9305$; here, $\overline{(\dots)}$ are
taken over $10^2$ different preparations. Dashed line: fitting law (\ref{3.3}) with $D=380$.
}
\end{figure}

\begin{figure}[t]
\includegraphics[angle=270]{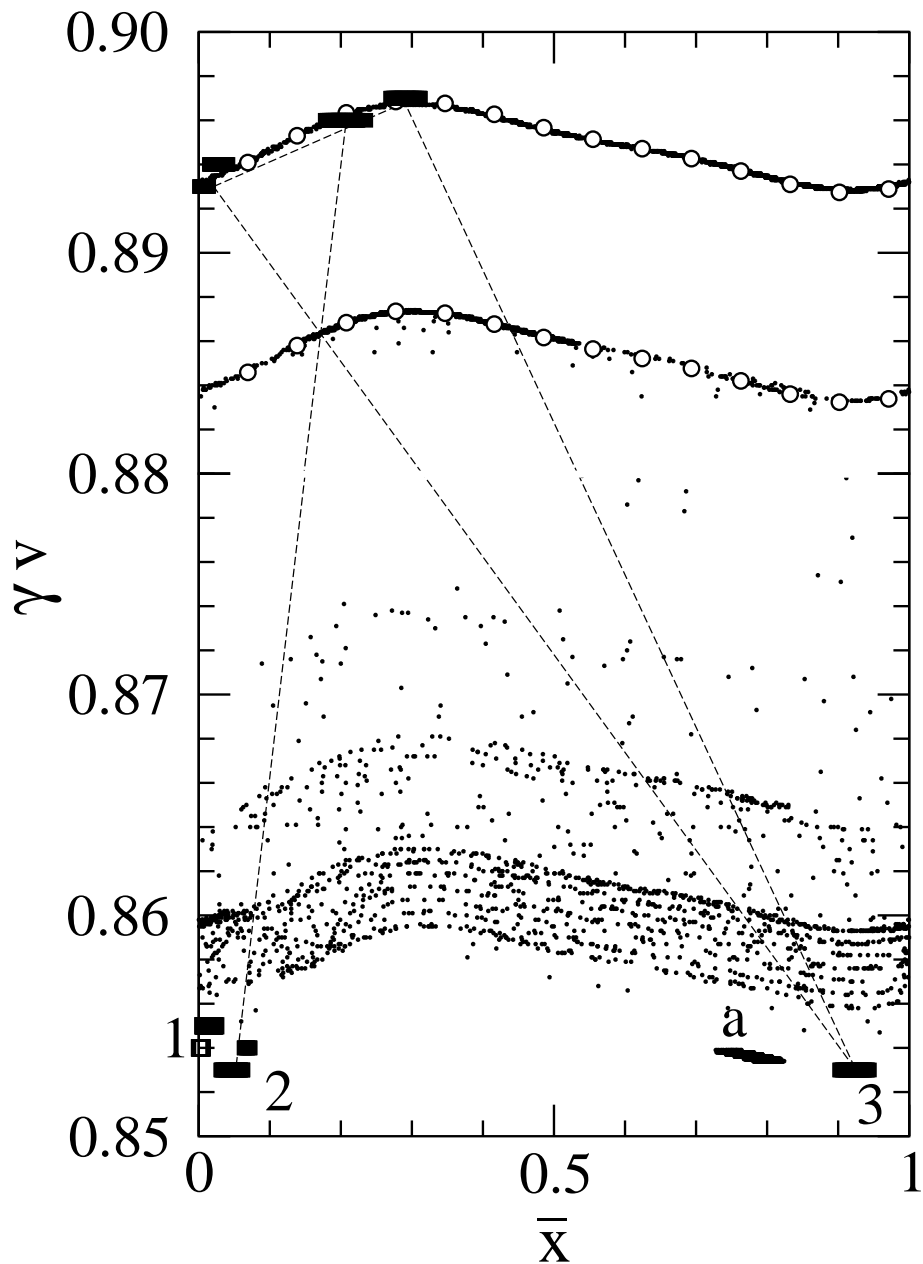}
\caption{\label{F4}
Reduced phase-space portrait for $A_0=0.9309$ (1), $0.9308$ (2),
 $0.9307$ (3), and  $0.9305$ (dots). Dashed straight lines connect the portrait point sets of (2) and (3).
Open circles:  $v_1(\overline{x})$ (upper) and $v_2(\overline{x})$ (lower), Eq. (\ref{4}), with $\overline{x_0}=-0.11$.
Note that the dots in the
diffusive regime are concentrated at the confined segment "a" and on the two upper bands.
The remaining  parameters are as in Fig. \ref{F3}.
}
\end{figure}
\end{widetext}

\end{document}